\documentclass[12pt]{article}
\usepackage{amssymb}
\usepackage{epsfig}

\newcommand{\bq}{\begin{equation}}
\newcommand{\eq}{\end{equation}}
\newcommand{\bqa}{\begin{eqnarray}}
\newcommand{\eqa}{\end{eqnarray}}
\newcommand{\ben}{\begin{enumerate}}
\newcommand{\een}{\end{enumerate}}
\newcommand{\bc}{\begin{center}}
\newcommand{\ec}{\end{center}}

\def\lsim{\lesssim}

\def\pr#1#2#3{ Phys. Rev. ${\bf{#1}}$ (#2) #3}
\def\prl#1#2#3{ Phys. Rev. Lett. ${\bf{#1}}$ (#2) #3}
\def\pl#1#2#3{ Phys. Lett. ${\bf{#1}}$ (#2) #3}

\def\np#1#2#3{ Nucl. Phys. ${\bf{#1}}$ (#2) #3}
\def\zp#1#2#3{ Z. f. Phys. ${\bf{#1}}$ (#2) #3}

\global\nulldelimiterspace = 0pt

\begin{document}
\thispagestyle{empty}
\begin {flushleft}

 PM/97-27\\

July 1997\\
\end{flushleft}

\vspace*{2cm}

\hspace*{-0.5cm}
\begin{center}
{\Large {\bf General constraints on light resonances}}\\ 
{\Large {\bf in a strongly
coupled symmetry breaking sector}}

\vspace{1.cm}
\centerline{\Large{R.S. Chivukula$^{\rm a,*,\dagger}$, F.M. Renard$^{\rm b}$,and 
            C. Verzegnassi$^{\rm c,*}$}}

\vspace {0.5cm} 
\centerline{\small  $^a$ \it Department of Physics, Boston University }
 \centerline{\small \it Boston, MA 02215, USA} 
\centerline{\small  $^b$ \it Physique
Math\'{e}matique et Th\'{e}orique, UPRES-A 5032,}
 \centerline{\small \it Universit\'{e} Montpellier
II, F-34095 Montpellier Cedex 5.}
\vskip 0.3cm
\centerline{\small $^{\rm c}$\it Dipartimento di Fisica, Universit\`a di 
                   Lecce and INFN, Sezione di Lecce,}
\centerline{\small \it via Arnesano, 73100 Lecce, Italy}

\vspace*{1cm}
{\bf Abstract}\hspace{2.2cm}\null
\end{center}
\hspace*{-1.2cm}
\begin{minipage}[b]{16cm}
In this paper we  consider the information that can be obtained about
a strongly-interacting symmetry breaking sector from 
precision measurements
of four-fermion processes at LEP2 or a (500 GeV) NLC . 
Using a ``Z-peak''
subtracted approach to describe four-fermion processes, we show that
measurements of the cross section for muon production, the related
forward-backward asymmetry, and the total cross section 
for the production
of hadrons (except $t$'s) can place constraints on 
(or measure the effects
of) the lightest vector or axial resonances present in a strong
symmetry breaking sector. We estimate that such effects 
will be visible at LEP2
for resonances of masses up to approximately 350 GeV, 
and at a 500 GeV NLC
for resonances of masses up to approximately 800 GeV. Multiscale models,
for example, predict the presence of light vector and axial mesons in 
this mass range and their effects could be probed by these measurements.
\\

\vspace{3cm}
(*) Work partially supported by NATO grant, CRG 951212\\
$^\dagger$ Work partially supported by DOE grant, DE-FG02-91ER40676\\
\end{minipage} 

\setcounter{footnote}{0} 
\clearpage
\newpage 
  
\hoffset=-1.46truecm
\voffset=-2.8truecm
\textwidth 16cm
\textheight 22cm
\setlength{\topmargin}{1.5cm}

\section{Introduction.}

\hspace{0.7cm}Precision measurements \cite{ewwg} of electroweak quantities at LEP and
SLC, as well as in neutral current and atomic parity violation
experiments, place constraints on models of dynamical electroweak
symmetry breaking. The most severe of these constraints arise from
measurements of the ``oblique" parameters S and T \cite{peskin,others},
which measure the extra isospin conserving and violating contributions
to the gauge boson self-energies, and from a measurement of the ratio of
widths $R_b=\Gamma_{Zb\bar b}/\Gamma_{Z had}$ \cite{rsczbb, zbbothers}.

The extra contributions to T and to $R_b$ in these models come largely
from the physics of top-quark mass generation. The absence of large
deviations in these quantities from their standard model predictions
strongly suggests that, if electroweak symmetry breaking occurs
dynamically, top-quark mass generation must take place in a separate
sector which {\it does not} give rise to the bulk of electroweak
symmetry breaking \cite{rscslac}. Examples of this kind include
top-condensate models \cite{topcondense}, such as top-color \cite{tc},
and top-color assisted technicolor \cite{tc2}. Contributions to $R_b$
and T can be largely eliminated, or at least greatly reduced, in models
of this sort. A natural implication in these cases is the existence of
extra light scalars, such as ``top-pions", and possibly light vector
resonances as well.

The constraints coming from S are potentially more problematic:
even if the bulk of electroweak symmetry breaking is due to 
an isospin-conserving strongly-interacting sector, there will
necessarily always be extra contributions to the gauge-boson
self-energies. Following Peskin and Takeuchi, an
estimate of the oblique effect S can be made as follows.
We model the contributions of the symmetry breaking
sector to the custodial triplet vector and axial
spectral functions each in terms of a {\it single
narrow} resonance:
\bq
        R_{VV,AA}(s) = 12 \pi^2 F^2_{V,A} \delta(s-M^2_{V,A})
\eq
The extra contribution from the symmetry breaking sector 
to $S$ can
then be written:
\bq
        S =4\pi\left[ {F^2_V\over M^2_V} - {F^2_A\over M^2_A}\right]~.
\eq
If we further assume that the electroweak symmetry breaking
sector is QCD-like, in that it obeys the analogs of the
first- and second-Weinberg sum rules \cite{weinberg}, we find that
this expression may be written
\bq
        S = 4\pi{F^2_\pi \over M^2_V}\left[1+{M^2_V\over M^2_A}\right]~.
\eq
where, to the extent that this sector is responsible for the bulk of
electroweak symmetry breaking, $F_\pi \approx 246 {\rm GeV}/\sqrt{N_D}$
is the analog of the pion decay constant in QCD.  Note that this
expression is always positive and, scaling from QCD, Peskin and Takeuchi
\cite{peskin} estimate it's value to be
\bq
        S \approx 0.25 {N_D N_{TC} \over 3}
\eq
where $N_D$ and $N_{TC}$ are the number of technidoublets
and technicolors respectively. This should be
compared with the most recent experimental value \cite{langacker}
\bq
        S = -0.26 \pm 0.16
\eq
some 3.5 standard deviations away from the estimate above.

As emphasized by Lane \cite{lane}, however, it is {\it necessary} that a
phenomenologically viable theory of dynamical electroweak symmetry
breaking be {\it unlike} QCD. As noted twenty years ago, in the absence
of a GIM mechanism {\it any} extension of a QCD-like symmetry breaking
sector that can account for the observed $s$ and $c$-quark masses is
likely to produce unacceptably large flavor-changing neutral-currents
\cite{fcnc}.  The most elegant mechanism proposed to solve this problem
is ``walking technicolor" \cite{walking}. In this case, unlike QCD, the
technicolor coupling does not fall {\it quickly} just above the chiral
symmetry breaking scale: i.e. the coupling ``walks" to zero instead of
running to zero quickly. This slowly falling coupling {\it enhances} the
masses of ordinary fermions and may allow one to construct theories
without dangerous flavor-changing neutral currents.

Part and parcel of this solution, however, is that
the asymptotic behavior of the coupling is {\it not}
like QCD \cite{lane}. This fact casts doubt on the use of
the first and second Weinberg sum rules in evaluating
S. Furthermore, some methods used to produce a walking
coupling invoke fermions in different representations
of the technicolor gauge group \cite{lanemulti}. In these ``multiscale"
models, TeV-scale resonances are expected from the bound-states
of the technifermions in the highest representation of the
gauge group, but lighter resonances occur as well. Indeed,
the lightest resonances may be as light as 300 -- 400 GeV
\cite{lanemulti}.

For these reasons, we will consider the implications of precision
electroweak data that can be gathered at future colliders
for more general strongly-interacting symmetry breaking sectors.
Instead of the simple, single-resonance
model used in \cite{peskin}, we consider the case when the
spectral function can be approximated as a {\it sum}
of narrow resonances, all beyond the production threshold:
\bq
        R_{VV}(s) = 12\pi^2 \sum_i F^2_{V_i} \delta(s-m^2_{V_i})
\eq
and
\bq
        R_{AA}(s) = 12\pi^2 \sum_j F^2_{A_j} \delta(s-m^2_{A_j})
\eq

The dispersive evaluation of S then gives:
\bq
        S=\left[\sum_i {F^2_{V_i}\over M^2_{V_i}} - 
        \sum_j {F^2_{A_j}\over M^2_{A_j}} \right]~.
\eq
In the absence of further information, very little can be determined
about the parameters in the expression above (note that, a priori, no
reason exists why the ratio $F^2_{V,A,n}/ M^2_{V,A,n}$ should
decrease \underline{quickly} with increasing n, even if the
resonances' masses do actually increase with n).

The aim of this paper is to show that precision measurements at high
energy $e^+ e^-$ colliders can conversely yield information about the
\underline{lightest} resonances in the vector and axial vector channel,
at least in a situation of ``reasonable" increase of the masses in the
families.
In the next section, we illustrate the method, in section 3, we perform
a quantitative analysis, and a short final discussion is given in
section 4.

\section{Section 2}
\subsection{Review of the approach}

\hspace{0.7cm}In a model that predicts the existence of one or several vector
resonances strongly coupled to the weak gauge bosons and to the photon,
one naturally expects that sizeable contributions to the gauge bosons
self-energies will arise from the transition 
gauge boson-resonance-gauge boson. In the
specific case of $e^+e^-$ annihilation, neutral resonances would
contribute , at the one loop level, the real part of the three
transverse self-energies $A_{\gamma}(q^2)$, $A_{\gamma Z}(q^2)$, 
$A_{Z}(q^2)$ where $q^2$ denotes the c.m. squared total energy of the
process and the definition of $A(q^2)$ is :

\bq  A_{i,j}(q^2)=A_{i,j}(0)+q^2 F_{i,j}(q^2) \label{2-1}
\eq
\noindent
($i,j=\gamma,Z$; $A_{i,i}\equiv A_i$). 

In fact, the situation is slightly more subtle at the one loop level.
In order to better understand this statement, it is useful to consider
the specific case of production of a final muon-antimuon couple, and in
particular to concentrate one's attention on the contribution to the
invariant scattering amplitude coming from the $Z$ exchange that, at
the tree level, reads

\bq T^{(0)}_{e\mu}(q^2,\theta)=i[{\sqrt2 G_{\mu,0}M^2_{Z,0}\over
q^2-M^2_{Z,0}}]\bar v_e\gamma^{\rho}(g^{(0)}_{Vl}-g^{(0)}_{Al}
\gamma^5)u_e\bar
u_{\mu}\gamma_{\rho}(g^{(0)}_{Vl}-g^{(0)}_{Al}\gamma^5)v_{\mu}
\label{2-2}
\eq

\noindent
where $G_{\mu,0}$, $M_{Z,0}$ are the ``bare" Fermi coupling and $Z$ mass
and we used the identity

\bq  \sqrt2 G_{\mu,0}M^2_{Z,0}={g^2_0\over4c^2_0}\label{2-3}
\eq
\noindent
($s^2_0=1-c^2_0={e^2_0\over g^2_0}$).

When one moves to one loop, it is straightforward to verify that the
term into the square bracket of eq.\ref{2-2} is replaced by

\bq  {\sqrt2 G_{\mu,0}M^2_{Z,0}\over
q^2-M^2_{Z,0}} \to {\sqrt2 G_{\mu}M^2_{Z}\over
q^2-M^2_{Z}+iM_Z\Gamma_Z}[1+({\delta G_{\mu}\over G_{\mu}}
+{Re \tilde {A}^{(e,\mu}_Z(0,\theta)\over
M^2_Z})-\tilde{I}^{e,\mu}_Z(q^2,\theta)]\label{2-4}
\eq
\noindent
where $G_{\mu}$, $M_{Z}$, $\Gamma_{Z}$ are now the physical
(conventionally defined) Fermi coupling, the $Z$ mass and the $Z$
width; $\tilde{A}_{Z}(q^2,\theta)$ is a certain
\underline{gauge-invariant} combination of self-energies, vertices and
boxes whose self-energy component is $A_{Z}(q^2)$ defined by eq.(1) and

\bq  \tilde{I}^{e,\mu}_Z(q^2,\theta)={q^2\over q^2-M^2_Z} 
Re[\tilde {F}^{(e,\mu}_Z(q^2,\theta)-{F}^{(e,\mu}_Z(M^2_Z,\theta)]
\label{2-5}\eq
\noindent
where the self-energy component of $\tilde{F}^{e\mu}_Z$ is
$F_{Z}(q^2)$, defined again by eq.(\ref{2-1}). The full expression of 
$\tilde{A}_{Z}(q^2,\theta)$ that includes vertices and boxes has been
given in previous references \cite{R16}. 
It has been obtained following closely
the Degrassi-Sirlin conventions and philosophy \cite{R17}, 
and it is
essentially based on the simple property that quantities contributing
different Lorentz structures of the invariant scattering amplitude must
be separately gauge-independent. Since the discussion of 
refs.\cite{R16} is
rather detailed, we shall not repeat it here, particularly because in
the special case that we are considering only the TC contributions to
the self-energies will be relevant. In this very definite sense, we can
from now on concentrate our attention on the transverse self-energy
content of the various gauge-independent quantities. The only possible
exception to this rule, provided by the process where a final $b\bar b$
couple is produced, will be separately discussed at the end of this
section.

Eq.(\ref{2-4}) shows the 
transformation at one loop of the term into the square
bracket of eq.(\ref{2-2}). 
To fully take into account the electroweak one loop structure it
is sufficient to replace the bare 
quantity $g^{0}_{Vl}$ in eq.(\ref{2-2})
by the corresponding finite and gauge-independent term

\bq  g^{0}_{Vl} \to g^{1}_{Vl} \equiv -{1\over2}[1-4s^2_l(q^2,\theta)]
\label{2-6}\eq
\noindent
where 

 \bq  s^2_l(q^2,\theta)=s^2_1[1+\tilde{\Delta}\kappa(q^2,\theta)]
\label{2-7}\eq

\bq  \tilde{\Delta}\kappa(q^2,\theta)={c_1\over s_1}[
\tilde {F}_{\gamma Z}(q^2)]+{c^2_1\over
c^2_1-s^2_1}[{\delta\alpha\over\alpha}-{\delta G_{\mu}\over G_{\mu}}-
{\delta M^2_Z\over M^2_Z}]
\label{2-8}\eq
\noindent
Here $\tilde {F}_{\gamma Z}(q^2)$ is a gauge-invariant combination 
of self-energies, vertices and boxes whose
self-energy part is $F_{\gamma Z}(q^2)$ defined by eq.(\ref{2-2})
and $s^2_1c^2_1\equiv s^2_1(1-s^2_1)={\pi\alpha
\over \sqrt2 G_{\mu}M^2_{Z}}$. Note that
one has at $q^2=M^2_Z$

\bq s^2_l(M^2_Z,\theta)\equiv  s^2_l(M^2_Z)
\eq
\noindent
(boxes do not contribute at this $q^2$ value) where $s^2_{l}(M^2_Z)$
is by definition the effective weak ``leptonic" angle measured on $Z$
resonance. Analogously, one finds

\bq  ({\delta G_{\mu}\over G_{\mu}}
+{Re \tilde {A}^{(e,\mu}_Z(0,\theta)\over
M^2_Z})-\tilde{I}^{e,\mu}_Z(M^2_Z,\theta)\equiv \epsilon_1
\label{2-9}\eq
\noindent
where $\epsilon_1$ is the first of the three Altarelli-Barbieri
parameters \cite{R18}, whose combination with the third parameter
$\epsilon_3$
produces $s^2_{l}(M^2_Z)$

\bq  s^2_{l}(M^2_Z)=s^2_1+{s^2_1\over
c^2_1-s^2_1}[\epsilon_3-c^2_1\epsilon_1+c^2_1(\Delta\alpha(M^2_Z)+
``vertices")]
\label{2-10}\eq
\noindent
where $\Delta\alpha(M^2_Z)=F_{\gamma}(0)-F_{\gamma}(M^2_Z)$
takes into account the running of $\alpha_{QED}$.\par
For the purpose of this paper, it is convenient to introduce now the
leptonic $Z$ width $\Gamma_l$, defined in terms of $\epsilon_{1,3}$
as

\bq  {\Gamma_l\over M_Z} =
{1\over24\pi\sqrt2}(1+\delta^{QED})G_{\mu}M^2_Z
[1+\epsilon_1][1+v^2_l(M^2_Z)]
\label{2-11}\eq
\noindent
where  $v_l\equiv1-4s^2_l$  and $\delta^{QED}$ is a known QED
``correction".\par
The previous expressions, that we wrote in order to achieve a fully
self-contained discussion, are sufficient to proceed now in a quicker
way towards our specific goal. The next step in this direction is that
of writing the expression for the ``Z-Z component" of the differential
$e-\mu$ cross section, that will read at one loop

\bq  ({d\sigma^{(1)(Z)}\over d\Omega})_{e\mu}(q^2,\theta)\simeq\
\{[{\sqrt2 G_{\mu}M^2_{Z}\over
q^2-M^2_{Z}}][1+({\delta G_{\mu}\over G_{\mu}}
+{Re \tilde {A}^{(e,\mu}_Z(0,\theta)\over
M^2_Z})-\tilde{I}^{e,\mu}_Z(q^2,\theta)]{1\over4}[1+(1-4s^2_l
(q^2,\theta))^2]\}^2
\label{2-12}\eq

The previous formula can be rewritten by formally 
``adding and subtracting" $\tilde{I}^{e,\mu}_Z(q^2=M^2_Z,\theta)$ and 
$s^2_l(q^2=M^2_Z,\theta)$. Throwing away higher order
terms and using the definitions eqs.(\ref{2-7}-\ref{2-12}) one concludes that
\underline{at the one loop level}, the alternative ``Z-peak subtracted"
representation can be used:

\bq ({d\sigma^{(1)(Z)}\over d\Omega})_{e\mu}(q^2,\theta)\simeq
\{[{\Gamma_l\over
M_Z}][1-R^{(l,\mu)}(q^2,\theta)-{8s_1c_1v_l(M^2_Z)\over1+v^2_l(M^2_Z)}
V^{(l,\mu)}(q^2,\theta)]\}^2
\label{2-13}\eq
\noindent
where the two ``form-factors"

\bq  R^{(e,\mu)}(q^2,\theta)\equiv \tilde{I}^{e,\mu}_Z(q^2,\theta)
-\tilde{I}^{e,\mu}_Z(M^2_Z,\theta)
\label{2-14}\eq
\noindent
and

\bq    V^{(e,\mu)}(q^2,\theta)\equiv
\tilde{F}^{e,\mu}_{\gamma Z}(q^2,\theta)
-\tilde{F}^{e,\mu}_{\gamma Z}(M^2_Z,\theta)
\label{2-15}\eq
\noindent
are now \underline{subtracted} at $q^2=M^2_Z$ and, in the remaining
theoretical expression, the Fermi coupling $G_{\mu}$ has disappeared,
having been replaced by quantities \underline{measured} on Z peak
($\Gamma_l$, $s^2_l(M^2_Z)$). Clearly, this will cause a loss of
precision in the theoretical prediction. This has been fully discussed
in Refs.\cite{R16}, showing that the accuracy of 
the experimental measurements
on $Z$ resonance is sufficient to guarantee that no sensible
theoretical error will be induced at LEP2, LC given their (optimal)
expected experimental accuracies. For this totally pragmatic reason, we
can conclude that eqs.(\ref{2-12}) and (\ref{2-13}) 
provide two practically equivalent
theoretical expressions for what concerns realistic future $e^+e^-$
colliders measurements in the two final muon channel.\par
In a totally analogous way, all the theoretical expressions for all
possible ``Z-Z" components of observables in the final two fermion
processes can be rewritten in such a way that the new expressions
contain Z-peak subtracted ``form-factors" and Z-peak measured
quantities (in general, Z widths and asymmetries). The full discussion
is given in refs.\cite{R16} and we do not insist on it here. 
Also, the method
can be easily extended to the various ``$\gamma-Z$" type components
leading to similar Z-peak subtracted expressions, while for the
``$\gamma-\gamma$" components the existing theoretical 
scheme does not need
any extra change: it will contain $\alpha_{QED}$ and ``$\gamma\gamma$"
form factors \underline{subtracted at $q^2=0$}, of the general 
\underline {non
universal} form

\bq \tilde{\Delta}\alpha^{(e,f)}(q^2,\theta)=\Delta\alpha(q^2)
+ ``l-f ~vertices,~ boxes"
\label{2-16}\eq
\noindent
where $\tilde{\Delta}\alpha^{(e,f)}(q^2=0,\theta)=0$, 
and the universal self-energy component $\Delta\alpha(q^2)$ is defined
after eq.(\ref{2-10}).\par
The next question that arises immediately at this point is that of
under which circumstances it will be convenient to abandon the
conventional representation where $G_{\mu}$ is retained and  to adopt
the Z-peak subtracted one. Although a general prescription is not
simple to be obtained, we can provide here an example of two specific
situations where, in our opinion, the use of the subtracted expression
will be rather useful. The first case is one in which one wants to write
down a simple, analytic theoretical expression, that may provide an
approximate ``satisfactory" estimate of the \underline{complete}
numerical value of various relevant observables in the Standard Model
at one loop. As we shall show in the next Section, this will indeed be
the case if a suitably defined ``Z-peak subtracted" Born approximation,
 built in the spirit of our approach, will be systematically utilized. 
The second case is one in which one wants to compute the virtual 
effects at one loop of a model of new physics that contains a number 
of parameters, whose $q^2$-dependence cannot be systematically
neglected, so that their effects in the subtracted form-factors will
not be generally negligible. Such would be the case of models of
technicolour type, to which we now devote the second part of this
section for a fully illustrative discussion.

\subsection{Effects of models of TC type}

\hspace{0.7cm}We shall start by considering the relatively simple case of a
technicolour-type model where two families of resonances
exist, one of vector and one of axial type, that can only contribute
self-energies through the chain gauge-boson-resonance-gauge-boson.\par
 Other possible effects
will be ignored. the only exception being represented by the
contribution to the $Zb\bar b$ vertex, to be discussed later on.
Therefore, assuming a number $N_V$ and a number $N_A$ of vector and
axial-vector resonances, under the reasonable assumption that the
various widths can be in first approximation neglected with respect to
the masses, the model can be parametrized by giving $2(N_V+N_A)$
parameters, conventionally called ``strengths" and masses, to be denoted
e.g. as $F^2_{Vi}$, $M_{Vi}$, $F^2_{Ai}$, $M_{Ai}$. Using as a first
approach a delta-type parametrization, we shall write following
conventional normalizations:

\bq  Im A_V(q^2)=\sum^{N_V}_{i=1} \pi q^2F^2_{Vi}\delta(q^2-M^2_{Vi})
\label{2-17}\eq

\bq  Im A_A(q^2)=\sum^{N_A}_{j=1} \pi q^2F^2_{Aj}\delta(q^2-M^2_{Aj})
\label{2-18}\eq
\noindent
where $A_{V,A}$ are the vector and axial-vector components of the three
neutral self-energies $A_{i,j}(q^2)$ eq.(\ref{2-1}), 
and we shall follow the
usual request that parity and strong isospin are conserved in the
model, writing

\bq  A_{\gamma\gamma}(q^2)=e^2A_V(q^2)
\label{2-19}\eq  

\bq  A_{ZZ}(q^2)={e^2\over4s^2_1c^2_1}[(1-2s^2_1)^2A_V(q^2)+A_A(q^2)]
\label{2-20}\eq

\bq 
A_{\gamma Z}(q^2)={e^2\over2s_1c_1}(1-2s^2_1)A_V(q^2)
\label{2-22}\eq
\noindent
(we are neglecting for the moment possible $\omega$-type
resonances).\par
The simplified delta function approximation of
eqs.(\ref{2-15}-\ref{2-18}) is very
useful provided that the use of \underline{unsubtracted} dispersion
relation is allowed to compute the resonant contribution to the
relevant self-energies. This provides a very general and to some extent
model independent way of proceeding. Whenever such an approach is not
feasible, other more specically model dependent approaches must be
used, that introduce in general extra parameters and assumptions.
Although this is in principle a tolerable attitude, the loss of
simplicity and generality thus introduced is certainly not an
advantage.\par
On this very specific point, the replacement of the conventional
representation by the Z-peak subtracted one plays the essential role.
This can be seen by concentrating again on the special example provided
by the equivalent eqs.(\ref{2-12}) and (\ref{2-13}). As one easily sees, the
self-energy content of eqs.(\ref{2-13}) is represented by three separate
quantities, i.e.
$({A_Z(0)\over M^2_Z}-{A_W(0)\over M^2_W})$, (the $W$ self-energy
$A_W(0)$
is coming from ${\delta G_{\mu}\over G_{\mu}}$), $I_Z(q^2)$ 
defined after eq.(5) and $\Delta\kappa(q^2)$ 
the self-energy component of eq.(8). Quite
generally, neither $({A_Z(0)\over M^2_Z}-{A_W(0)\over M^2_W})$ 
nor $\Delta\kappa(q^2)$ satisfy unsubtracted dispersion
relations, since in the limit of large $q^2$ all $F_i(q^2)$ approach a
non vanishing constant (in particular, $A_i(q^2)$ diverges linearly).
This difficulty is completely removed in eq.(\ref{2-13}), where only
\underline{subtracted} quantities $\simeq[F_i(q^2)-F_i(M^2_Z)]$ 
survive, since for such
differences it is always possible to write the (formally)
\underline{unsubtracted} dispersion relation:

\bq  [F_i(q^2)-F_i(M^2_Z)]=({q^2-M^2_Z\over\pi})P\int^{\inf}_0 {ds
Im F_i(s)\over(s-q^2)(s-M^2_Z)}
\label{2-23}\eq

As a consequence of this fact, the complete TC resonant contribution to
self-energies in LEP2, LC two fermion production in the approximation
of eqs.(\ref{2-18}),(\ref{2-19}), 
will be expressed in terms of the resonances'
strengths and masses, and more precisely it will affect the
three universal
form-factors in the following way:

\bqa  R^{(TC)}(q^2)=&&{\pi\alpha(q^2-M^2_Z)\over
s^2_1c^2_1}\{(1-2s^2_1)^2\sum_i {F^2_{Vi}\over
M^2_{Vi}}({1\over1-{M^2_Z\over M^2_{Vi}}})^2({1\over
M^2_{Vi}-q^2})+\nonumber\\
&&\sum_i {F^2_{Ai}\over
M^2_{Ai}}({1\over1-{M^2_Z\over M^2_{Ai}}})^2({1\over M^2_{Ai}-q^2})\}
\label{2-24}\eqa

\bq V^{(TC)}(q^2)={2\pi\alpha(q^2-M^2_Z)\over
s_1c_1}\{(1-2s^2_1)\sum_i {F^2_{Vi}\over M^2_{Vi}}
({1\over1-{M^2_Z\over M^2_{Vi}}})({1\over M^2_{Vi}-q^2})\}
\label{2-25}\eq

\bq \Delta^{(TC)}(q^2)=-4\pi\alpha\sum_i 
{F^2_{Vi}\over M^2_{Vi}}({1\over M^2_{Vi}-q^2})
\label{2-26}\eq

Eqs.(\ref{2-24}-\ref{2-26}) can be easily 
derived starting from the definitions of
$R$, $V$, $\Delta\alpha$ eqs.(\ref{2-15}), (\ref{2-16}), (\ref{2-11}) 
and from eqs.(\ref{2-18}-\ref{2-22}).
Note that the delta-function approximation is not essential in our
derivation. More realistic parametrizations {\it a-la} 
Breit-Wigner could be
used, but for not unreasonably large resonance widths the results would
not be essentially modified, as one could easily verify.\par
Starting from eqs.(\ref{2-24}-\ref{2-26}) 
it is now straightforward to compute the TC
contribution to all the observables of the process $e^+e^-\to f\bar f$
at variable $q^2$. This will be discussed in detail in the next final
Section 3. But before doing that we want to stress another consequence
of the use of our subtracted approach that is, we believe, rather
useful. Looking at eqs.(\ref{2-24},\ref{2-25}) 
one sees that in the sums over the
two family indices each separate $F^2_i/M^2_i$ term is multiplied by a
factor $\simeq 1/(M^2_i-q^2)$. This has the consequence that a
factor $\simeq 1/ [1+(M^2_n-M^2_1)/( M^2_1-q^2)]$ 
will weight the $n$-th coefficient ${F^2_n\over M^2_n}$ in the sum, thus
strongly depressing the $n>1$ contributions when $q^2$ increases (and
approaches values not dramatically smaller than $M^2_1$) and $M_n$,
$n>1$, is ``reasonably" larger than $M_1$. This
mechanism favours therefore the possibility of isolating, or at least
priviledging, the contributions coming from the lightest vector and
axial-vector resonances, independently of the existence and the
features of heavier components in the family.\par
This situation should be compared with that obtainable from
measurements on Z resonance, that was exhaustively discussed in
refs.\cite{peskin}. In particular, 
the effect of a TC model like the one that
we are considering here on $\epsilon_3$ (which is equivalent to the
effect on the self-energy part of $\epsilon_3$, practically on the
Peskin-Takeuchi parameter $S$) would be of the following kind:

\bq  \epsilon^{(TC)}_3 \simeq {\alpha\over4s^2_1}S^{TC}\simeq 
{\alpha\over s^2_1}\{\sum_i {F^2_{Vi}\over
M^2_{Vi}}-\sum_j {F^2_{Aj}\over
M^2_{Aj}}\}
\label{2-27}\eq
\noindent
(we did note impose in eq.\ref{2-2} the validity of the two Weinberg sum
rules).\par
A glance at eq.(\ref{2-27}) shows that, indeed, the type of information
derivable at LEP1 from $\epsilon_3$ ($S$) on a model with families is
quite different and therefore independent of that derivable at larger
$q^2$ values at LEP2, LC. In fact, no depression factor exists in
eq.(\ref{2-27}), and one measures in fact the \underline{sum} of all 
${F^2_{i,j}\over
M^2_{i,j}}$
terms. Only for the special case of one resonance per family, the
informations at LEP1, LEP2, LC are of the same type.\par
Keeping this fact in mind, in the remaining part of this paper we shall
examine a case where only the 2 lightest resonances 
(a ``techni-$\rho$" and
a ``techni-$A$") will contribute  eqs.(\ref{2-24}-\ref{2-26}). 
In other words, we shall
assume that the next terms in the two sums are sufficiently depressed
with respect to the lightest ones. The quantitative analysis of this
particularly simple case will be shown in the forthcoming Section 3.

\section{TC contribution to realistic observables}

\subsection{Formal expressions}

\hspace{0.7cm}From the expressions of the TC contribution to the three universal form
factors $R$, $V$, $\Delta\alpha$ given by eqs.(\ref{2-24}-\ref{2-26}) 
it is simple to
derive the corresponding effect on \underline{any} observable quantity
of the process $e^+e^-\to 2$ fermions. In practice, though, this
procedure is unlikely to lead to interesting consequences unless the
predicted effect is, on a certain suitable scale, relatively ``large".
In particular, a rather essential condition is that the experimental
accuracy with which a certain observable can be measured is ``suitable",
since the size of ``visible" effects should be in any case larger ( by
an amount dictated by the aimed confidence level of the research) than
that of the experimental error.\par
At LEP2, at the optimal expected experimental conditions illustrated in
previous studies \cite{R19}, the best experimental precision, 
of a relative
size of less than one percent, will be achieved in the measurements of
\underline{three} observables i.e. $\sigma_{\mu}$ (the cross section
for muon production), $A_{FB,\mu}$ (the related forward-backward
asymmetry) and $\sigma_5$ (the cross section for production of
hadrons). The same situation will probably occur at a next linear
$e^+e^-$ collider of 500 GeV (NLC) \cite{R20} 
at least in a first stage where
initial state longitudinal polarization will not be achievable. For
this reason, we have decided in this paper to concentrate our attention
on the three previously considered observables, leaving a more
complete analysis of extra quantities (like specific heavy quark
production cross sections and asymmetries and/or various longitudinal
polarization asymmetries) to a future investigation.\par
 As a first step towards our goal, we will write three approximate
expressions \underline{valid} \underline{at one loop}, 
where only the self-energy
component of the various form factors has been retained. In order to
avoid unnecessary complications, we shall moreover retain only, for
what concerns the SM component of the involved self-energies, the pure
fermionic contribution (the latter is automatically gauge-invariant).
Also, terms that are numerically irrelevant (at the percent level fixed
by the experimental accuracy) will be systematically neglected,
following the discussion given in refs.\cite{R16}. As a result of this
procedure, we shall be led to the following approximate expressions,
valid in the ``Z-peak subracted" approach (and, strictly, at the one
loop level):

\bqa \sigma_{\mu}(q^2)\simeq &&{4\over3}\pi q^2
\{[{\alpha^2\over q^4}][{1\over1-\Delta\alpha(q^2)}]^2
+[{3\Gamma_l\over M_Z}({1\over q^2-M^2_Z})]^2
[1-R(q^2)-8s_1c_1v_l(M^2_Z)V(q^2)]^2\nonumber\\
&&+[({2\alpha\over q^2})
({3\Gamma_l\over M_Z})({1\over
q^2-M^2_Z})v^2_l(M^2_Z)]\}
\label{2-28}\eqa

A few comments an eq.(\ref{2-28}) are appropriate at this point. 
The contribution
from the ``Z-Z" term can be reconstructed directly from eq.(\ref{2-14}). 
For
sufficiently large $q^2$ values, $q^2>M^2_Z$, it is much less important
than the pure photon contribution. For the latter we have retained a
factor $1/(1-\Delta\alpha)$ that, for what concerns its SM fermionic component, resums
over all the leading logarithmic contributions. For contributions from
other sources, i.e. of TC origin, we shall identify (at our one-loop
level) $1/(1-\Delta\alpha^{(TC)})$ with
$[1+\Delta\alpha^{(TC)}]$.\par
The third contribution in eq.(28) is due to $\gamma-Z$ interference.
This is, in the case of $\sigma_{\mu}$, definitely smaller than the
other two terms (almost two orders of magnitude) at lowest order. We
have not used in this case the extra $O(\alpha)$ correction factor,
that would contain $\Delta\alpha$, $R$ and $V$, since the practical
effect of this correction would be invisible, at the given experimental
conditions.\par
The derivation of eq.(\ref{2-28}) is particularly simple. In a perfectly
equivalent way one can derive two approximate expressions for the
remaining considered onservables, as it has been shown in
refs.\cite{R16}
More precisely, we shall define :

\bq  A_{FB,\mu}(q^2)=[{\sigma_{FB,\mu}(q^2)\over\sigma_{\mu}(q^2)}]
\label{2-29}\eq
\noindent
where $\sigma_{\mu}$ will be approximated by eq.(\ref{2-28}) and

\bq  \sigma_{FB,\mu}(q^2)\simeq \pi q^2\{({2\alpha\over
q^2})({3\Gamma_l\over M_Z})({1\over
q^2-M^2_Z})({1\over1+v^2_l(M^2_Z)})[{1-R(q^2)\over1-\Delta\alpha(q^2)}]+
({3\Gamma_l\over M_Z})^2{4v^2_l(M^2_Z)\over(q^2-M^2_Z)^2}]\}
\label{2-30}\eq

Note that, away from the Z peak, the pure Z contribution (second term
in eq.(\ref{2-30}) is negligible with respect to the $\gamma-Z$ interference
(first term). For this reason, like in the previous case, we did not
consider the extra correction factor containing $R$ and $V$ that would
multiply it.\par
Quite similar considerations, that lead to several welcome
simplifications, finally allow to write an approximate expression
for $\sigma_5(q^2)$ ($\equiv
\sigma_u(q^2)+\sigma_d(q^2)+\sigma_s(q^2)+\sigma_c(q^2)
+\sigma_b(q^2)$) 
that is relatively simple, and reads:

\bqa  \sigma_5(q^2)\simeq &&N^{QCD}_q(q^2){4\over3}
\pi q^2\{[{11\alpha^2\over
9q^4}][{1\over1-\Delta\alpha(q^2)}]^2+\nonumber\\
&&[({3\Gamma_l\over M_Z})({3\Gamma_{had}\over
M_Z N^{QCD}_q(M^2_Z)})({1\over
q^2-M^2_Z})][1-2R(q^2)-s_1c_1V(q^2)({32\over10}{
\Gamma_c\over\Gamma_{had}}+{48\over13}{\Gamma_b\over\Gamma_{had}})]\}
\nonumber\\
&&+[({2\alpha\over q^2})({2v_l(M^2_Z)\over
q^2-M^2_Z})\sqrt{({3\Gamma_l\over M_Z})({3\Gamma_{had}\over
M_ZN^{QCD}_q(M^2_Z)})}({2\over9}\sqrt{{
\Gamma_c\over\Gamma_{had}}}+{3\over\sqrt3}
\sqrt{{\Gamma_b\over\Gamma_{had}}})\}
\label{2-31}\eqa
\noindent
where $\Gamma_c$, $\Gamma_b$, $\Gamma_{had}$ are the Z widths into 
$c\bar c$, $b\bar b$ and hadrons (measured on top of Z
resonance) and, to obtain some of the numerical factors, the
appropriate expression $s^2_l=1/4$ has been used (only in the terms
that are already negligibly small or of $O(\alpha)$. Note that, in the
case of $\sigma_5$, the pure photon and the pure Z terms are almost
equivalent, and much more important than the $\gamma-Z$ interference
contribution (last term in eq.(\ref{2-31})), 
for which consequently several
approximations (like that of considering $\Gamma_u=\Gamma_c$ and 
$\Gamma_d=\Gamma_s=\Gamma_b$)
have been used, whose effect on the overall expressions is well beyond
the experimental accuracy, as fully discussed in ref.\cite{R16}.\par
From eqs.(\ref{2-28}-\ref{2-31}) it is straightforward 
to derive the TC vector
resonances (VR) virtual contribution (at one loop) to the considered
observables. More precisely, we shall have in our approach the
following TC ``shifts":

\bqa \delta\sigma^{(TC)VR}_{\mu}(q^2)=&&{4\over3}\pi q^2
\{[{\alpha^2\over
q^4}]2\Delta\alpha^{(TC)}(q^2)\nonumber\\
&&-[({3\Gamma_l\over M_Z})({1\over
q^2-M^2_Z})]^2[2R^{(TC)}(q^2)+16s_1c_1v_l(M^2_Z)V^{(TC)}(q^2)]\}
\label{2-32}\eqa

\bqa \delta\sigma^{(TC)VR}_{FB,\mu}(q^2)=&&N^{QCD}_q(q^2){4\over3}\pi q^2
\{[{11\over9}{\alpha^2\over
q^2}]2\Delta\alpha^{(TC)}(q^2)\nonumber\\
&&-[({2\alpha\over q^2})
({3\Gamma_l\over M_Z})({1\over
q^2-M^2_Z})({1\over1+v^2_l(M^2_Z)})][\Delta\alpha^{(TC)}(q^2)
-R^{(TC)}(q^2)]\}
\label{2-33}\eqa

\bqa \delta\sigma^{(TC)VR}_5(q^2)=&&{4\over3}\pi q^2
\{[{\alpha^2\over q^4}]2\Delta\alpha^{(TC)}(q^2)
-[({3\Gamma_l\over M_Z})({3\Gamma_{had}\over M_Z}({1\over
q^2-M^2_Z})^2]\nonumber\\
&&[2R^{(TC)}(q^2)+s_1c_1V^{(TC)}(q^2)({32\over10}{
\Gamma_c\over\Gamma_{had}}+{48\over13}{\Gamma_b\over\Gamma_{had}})]\}
\label{2-34}\eqa

Starting from eqs.(\ref{2-32}-\ref{2-34}) and taking into account the definition 
eq.(\ref{2-29}) one can easily obtain the formal expressions of the various
effects by simply inserting eqs.(\ref{2-24}-\ref{2-26}). At this point, everything
is ready for the derivation of exclusion limits on the parameters 
from suitable conventional fits, at a given confidence level. This will
be done in the forthcoming Section 3.2 . But before doing that we want
to devote the last part of this half-Section to a quantitative
discussion, related to the actual accuracy of the approximated formulae
eqs.(\ref{2-28})-\ref{2-31}) that we have used for what concerns the
\underline{conventional Standard Model prediction}.\par
On the basis of the ``Z-peak subtracted" philosophy, we would actually
expect that eqs.(\ref{2-28})-\ref{2-31}) 
should give a reasonable approximation,
within the SM scheme, of all those contributions that have been already
``partially'' reabsorbed by quantities measured either at $q^2=M^2_z$ (Z
widths, asymmetries) or at $q^2=0$ ($\alpha$). The genuinely
electroweak terms that cannot be reabsorbed in this way (QED boxes are
supposedly known and separately calculable) are those coming from weak
boxes, since these are kinematically vanishing at these points. This
suggests that a more ambitious approximation to one-loop observables
should contain the effect from weak box diagrams, in particular at
higher $q^2$ values where such terms might become more relevant (e.g.
``pure Z" boxes will be multiplied in our representation by $(q^2-M^2_Z)$
factors). In fact, the complete existing programs \cite{R21}
 confirm this
feeling in details.\par
In order to achieve this additional accuracy, we have proceeded as
follows. First of all, we have decided from now on to work for what
concerns the SM component in the t'Hooft $\xi=1$ gauge. In this gauge,
we have then simply computed the SM WW and ZZ boxes following the
prescriptions given by Consoli and Hollik \cite{R22}. 
These extra terms have
then been added to the SM component of eqs.(\ref{2-28}-\ref{2-31}). 
The resulting
expressions are those that we expect to provide a satisfactory
approximation to the SM predictions for $q^2>M^2_Z$. They contain a
certain amount of vertices already reabsorbed in the new theoretical
inputs; the SM self-energy content should be practically negligible (it
is systematically subtracted) with the exception of the
fermionic contribution to $\Delta\alpha(q^2)$, 
that is in fact the only quantity that
we have retained; the weak boxes content is exactly added. We have
defined these approximate expressions with an ``S" index, that means
``subtracted", and written formally

\bq  \sigma^{(SM,S)}_{\mu}(q^2)\equiv [\sigma_{\mu}(eq.(28))]^{SM}
+ \sigma^{W,Z boxes}_{\mu}(q^2)
\label{2-35}\eq

\bq  A^{(SM,S)}_{FB,\mu}(q^2)\equiv [A_{FB,\mu}(eq.(29-30))]^{SM}
+ A^{W,Z boxes}_{FB,\mu}(q^2)
\label{2-36}\eq

\bq  \sigma^{(SM,S)}_5(q^2)\equiv [\sigma_5(eq.(31))]^{SM}
+ \sigma^{W,Z boxes}_5(q^2)
\label{2-37}\eq

To verify whether our feelings were correct, we have then carried on a
systematic comparison with the rigourous predictions obtainable e.g. by
using the available program TOPAZ0 \cite{R21} .
We show the results of this
comparison in the following Table I at several $q^2$ values, chosen for
simplicity in the LEP2 range i.e. for $\sqrt{q^2}\lsim 200~GeV$.
Note that the purely electroweak contribution to $\sigma_5$ has been
computed. In order to obtain the complete expression that includes
the strong interaction effects, a conventional rescaling factor
$(1+\alpha_s(q^2)/\pi)$ should be applied. This is straightforward and
should not require any special comment. \par
As one sees from inspection of Table I, the numerical agreement between
our approximate ``subtracted" (S) expressions
eqs.(\ref{2-35}-\ref{2-37}) 
and the full
rigorous TOPAZ0 (T) calculation is indeed impressive, since the
numerical differences are systematically well below the optimal
expected ($\simeq 1 \%$) experimental accuracy, with one possible
exception around the \underline{critical} value $\sqrt{q^2}=2M_W$,
where the ``resonant" effects of the W vertex, that cannot be taken into
account by our method, are not negligible. This means that, for
$\sqrt{q^2}$ not very close to $2M_W$, we shall be entitled to use our
expressions as an adequate description of the SM predictions
\underline{as} long as searches of visible effects of New Physics
will be performed. This is exactly what we shall do in the next
forthcoming Section 3.2.

\subsection{Numerical analysis of the TC effects}

\hspace{0.7cm}From the previous long discussion we have concluded that there might be
an effect in the three observables $\sigma_{\mu}$, $A_{FB,\mu}$ and
$\sigma_5$, at $q^2>M^2_Z$, essentially produced by the two
lightest members of two families of vector and axial vector resonances,
when the next resonances are sufficiently heavier than the lightest
ones, that are in turn supposed to be not too far away from the
available total c.m. $e^+e^-$ energy. In this case, there would be four
parameters to be fitted at a given $q^2$ i.e. ${F^2_V\over M^2_V}$,
$M^2_V$, ${F^2_A\over M^2_A}$, and $M^2_A$ (from
now on $V$, $A$ will denote the lightest resonances).\par
We can introduce at this point another simplification that is somehow
motivated by technical considerations. More precisely, we shall accept
that both $M^2_V$, $M^2_A$ are definitely larger than $M^2_Z$:

\bq  {M^2_Z\over M^2_V} << 1 \ \ \ ;\ \ \  {M^2_Z\over M^2_A} << 1
\label{2-38}\eq

The practical consequence of this assumption is that the number of
parameters that appear in the three form-factors $\Delta\alpha$, $R$,
$V$ eqs.(\ref{2-24}-\ref{2-26}) is now reduced to two, i.e.

\bq  X(q^2)= ({F^2_V\over M^2_V})({1\over M^2_V-q^2})
\label{2-39}\eq

\bq  Y(q^2)= ({F^2_a\over M^2_A})({1\over M^2_A-q^2})
\label{2-40}\eq
\noindent
and the numerical program devised to calculate bounds at a given
confidence level becomes rather simple. We have decided to proceed in
this way to get a first set of results valid in the approximation of
eq.(\ref{2-38}), encouraged by the fact that our ansatz is actually only
affecting the two form factors $R$ and $V$ but not $\Delta\alpha$ that,
in the three considered observables, gives a large part of the effect
(in the particular case of $\sigma_{\mu}$, $\Delta\alpha$ is in
practice the only relevant correction). Therefore our approximation
will only affect a fraction of the calculation, mostly for what
concerns the contribution of the axial vector resonance (and only if
$M^2_A$ is not sufficiently larger than $M^2_Z$).\par
In the first part of our analysis we have derived model-independent
bounds for the two quantities $X(q^2)$, $Y(q^2)$ at two different
values of $\sqrt{q^2}=192~GeV$ and $500~GeV$ that correspond to the
highest energy situations at LEP2 and at a future 
NLC $e^+e^-$ linear collider. With
this aim, we have parametrized the \underline{relative} TC shifts of
the considered observables from their SM values as follows 

\bq  {\delta\sigma^{(TC)}_{\mu}\over \sigma_{\mu}}(q^2)=
{\sigma^{(TC)}_{\mu}-\sigma^{(SM)}_{\mu}\over \sigma_{\mu}}=
a_1(q^2) X(q^2)+b_1(q^2) Y(q^2)
\label{2-41}\eq

\bq   {\delta A^{(TC)}_{FB,\mu}\over A_{FB,\mu}}(q^2)=
{A^{(TC)}_{FB,\mu}-A^{(SM)}_{FB,\mu}\over A_{FB,\mu}}=
a_2(q^2) X(q^2)+b_2(q^2) Y(q^2)
\label{2-42}\eq

\bq   {\delta\sigma^{(TC)}_5\over \sigma_5}(q^2)=
{\sigma^{(TC)}_5-\sigma^{(SM)}_5\over \sigma_5}=
a_3(q^2) X(q^2)+b_3(q^2) Y(q^2)
\label{2-43}\eq
\noindent
where the quantities $a_j$, $b_j$ are certain functions of $q^2$ whose
numerical values can be easily derived from eqs.(\ref{2-32},\ref{2-34}) 
and from the
given expressions of the SM formulae used in our program.\par
In the practical derivation of bounds we have assumed that the
experimental accuracies for the three observables are, both at LEP2
\cite{R19}
and at NLC \cite{R20}, respectively 

\bq {\delta\sigma^{(exp)}_{\mu}\over \sigma_{\mu}}=
{\delta A^{(exp)}_{FB,\mu}\over A_{FB,\mu}}=
\pm0.009;\ \ \ \ \ 
{\delta\sigma^{(exp)}_5\over \sigma_5}=\pm0.007
\label{2-44}\eq

Under the assumption of non observability of any deviation from the SM
predictions, we have then derived the $95\%$ CL exclusion curves in
$TeV^{-2}$ for
$X$, $Y$ depicted in Fig.(1) (LEP2) and Fig.(2) (NLC) 
respectively (since
both $X$ and $Y$ are by definition positive quantities, we have only
shown the relevant quadrant in the ($X$, $Y$) plane). As one sees, the
dependence of $X$ and $Y$ is rather different in the observables: in
particular, $\sigma_{\mu}$ gives, as expected, essentially information
on $X$, while $\sigma_5$ and the asymmetry are also sensitive to $Y$.
As a consequence of the different sensitivities, the combined $95\%$ CL
exclusion ellipse can be reasonably approximated in the allowed ($X$,
$Y$ $>0$) region, by a line of equation

\bq  Y=a+b X 
\label{2-45}\eq
\noindent
with $(a,b)=(2.8, -1.7)$ at LEP2 and $(0.38, -1.8)$ at NLC.
In practice, the limiting values
would be roughly of $O(1)$ at LEP2 and ten times smaller at NLC.\par
Figs.(1) and (2) show a result that is, to a certain extent, model
independent in the sense that no specific assumption e.g. on the
strengths of the couplings has been used. From the curves given in
these Figures one can, as a next step, derive bounds for a subset of
the four original parameters ($F^2_i/M^2_i, ~M^2_i$) if certain conditions on the
remaining subset are imposed. To give an example of such a procedure we
have first parametrized the two effective strengths in the following
way:

\bq  {F_V\over M_V}=\alpha {f_{\rho}\over m_{\rho}}\ \ ;\ \ \ 
{F_A\over M_A}=\beta{F_V\over M_V} 
\label{2-46}\eq
\noindent
where $f_{\rho}$ and $m_{\rho}$ are the $\rho$ parameters and 
${f_{\rho}\over m_{\rho}}\simeq 1/(2\sqrt{2\pi})$.
The choice $\alpha=1$ would be dictated by QCD analogy. From our
previous introductory considerations, we are led to discard values of
$\alpha$ too close to $1$. On the other hand, smaller $\alpha$ values
would lead to unduely pessimistic boundaries values for the resonant
masses. To proceed with a mildly optimistic compromise, we chose
therefore the value $\alpha=2$. For a first investigation we also fixed
the values $\beta=1$ and redid our analysis having now as residual free
parameters the two masses $M_V$ and $M_A$.\par
The results of our investigation for this particular case are shown in
Figs.(3) (LEP2) and (4) (NLC). One sees that the $95\%$ CL exclusion
limits would now lie in a range of approximately 
$M_V\simeq500~GeV$, $M_A\simeq350~GeV$
for LEP2(192) 
and $M_V=1.5~TeV$, $M_A=0.9~TeV$ for NLC(500). 
Although these values are given for a special choice
of the strengths, they provide a qualitative measure of what could be a
realistic reach for searches of two lightest technivectors in models of
the considered type, with reasonable and self-consistent choices of the
two effective strengths ${F^2_V/ M^2_V}$ and ${F^2_A/ M^2_A}$.
 Clearly, from the general result
given in Figs.(1) and (2), one would easily derive other types of
bounds for different input choices dictated by the specific model.\par
Having established, at least qualitatively, the type of negative bounds
that will be derivable at LEP2, NLC, we tried to examine the more
optimistic possibility of indentification of a TC signal.\par
Let us first notice that from eq.(\ref{2-41}-\ref{2-43}) a TC signal
should satisfy the general constraint (independent of the precise
values of the TC parameters):

\bq {\delta A^{(TC)}_{FB,\mu}\over A_{FB,\mu}}\simeq
c_1{\delta\sigma^{(TC)}_5\over \sigma_5}  
+c_2{\delta\sigma^{(TC)}_{\mu}\over \sigma_{\mu}}
\label{2-47}\eq
\noindent
with ($c_1\simeq 0.9, ~c_2\simeq -1.2$) at LEP2 and
($c_1\simeq 1.3, ~c_2\simeq -1.6$)  at NLC.\\
From eq.(\ref{2-47}), it would not be difficult to draw the plane that,
in the space of the three shifts, would correspond to the considered TC
model. For simplicity, at this qualitative
stage,
rather than one three-dimensional plot,
we considered three two-dimensional ones in the planes $(\sigma_{\mu},
~A_{FB,\mu})$, 
$(\sigma_{\mu},~\sigma_5)$ 
and $(A_{FB,\mu},~\sigma_5)$,
respectively. The SM values were computed using our ``subtracted"
program, and the TC contributions was added following our previous
equations, without computing in a first stage the QED radiation
effects.\par
In Figs.(5),(6) we show a typical situation occuring in the 
$(\sigma_{\mu},~\sigma_5)$ plane. This
is in fact the most interesting situation, since these two variables
are more sensitive than the asymmetry to the TC effects. Moreover,
there is in this case a rather typical feature represented by the fact
that the TC shifts on $\sigma_{\mu}$ and $\sigma_5$ 
are essentially \underline{negative}.
This is true whenever the parametrization eq.(\ref{2-24}-\ref{2-26}) 
for $R^{(TC)}$, $\Delta\alpha^{(TC)}$
can be used, independently of the number of resonances, 
and would still be valid if a
less simplified parametrization a la Breit-Wigner were used. To
conclude that a signal of possible TC origin was seen, one should first
verify whether the shift lies in the ``suggested" region in the 
$(\sigma_{\mu},~\sigma_5)$
plane, and consider this as a (quasi) necessary condition. In order to
make more definite claims, though, a derivation of $\simeq 5\sigma$
should be found. This would require values of the masses, in our
representative example, of order ($M_V\simeq 310~GeV,
~M_A\simeq310 ~GeV$) or ($M_V\simeq 470~GeV,
~M_A\simeq260~ GeV$) at LEP2. 
In the
more ambitious situation of NLC, one would be able to see an effect for
($M_V\simeq 0.8~TeV,
~M_A\simeq0.8 ~TeV$) or ($M_V\simeq 1.4~TeV,
~M_A\simeq 0.67~ TeV$)
. \par
The previous conclusions were drawn without taking into account the QED
radiation effects. The reason why we proceeded in this way is that such
an analysis was already performed in a previous paper \cite{R23} for a
situation that is extremely similar to that met in this paper. In ref
\cite{R23},in fact, the QED convoluted effect was computed using the so-called
structure-functions approach \cite{R24} for a case in which the SM prediction,
computed essentially in the same ``subtracted" philosophy, was
implemented by corrections due to assumed anomalous triple gauge boson
couplings. These were parametrized by simple analytic expressions that
were certain functions of $q^2$ only (and not of the scattering angle
$\theta$). The conclusion of ref \cite{R23} was that, provided that a suitable
cut on the hard photon energy was enforced, the simple ``not QED
convoluted" effects were totally unaltered. Since our TC model affects
the SM predictions by a simple $q^2$ dependent function, the
conclusions of ref \cite{R23} will remain the same, as we checked in a few
simple representative cases. We do not insist on this point here, since
it has been rather exhaustively discussed in the aforementioned
ref.\cite{R23}
to which we defer  the interested reader for more details. We simply
say that Fig.(5) can be considered as a realistic one, that is not
altered by the QED radiation effects which, for $\sigma_{\mu}$,
$A_{FB,\mu}$ and $\sigma_5$, can be eliminated
by a cut of ``standard" type.\par
To conclude this Section, we would like to add a final comment about
possible TC effects in the $Zb\bar b$ vertex 
that we have not considered in
our analysis. These would not affect $\sigma_{\mu}$ and 
$A_{FB,\mu}$ and would enter $\sigma_5$
since they could affect the $b\bar b$ component $\sigma_b$.\par
Two comments are useful at this point. The first one is that, in any
case, only the ``pure Z" component of $\sigma_b$ in $\sigma_5$ would
be modified via $Zb\bar b$ vertex effects. Numerically, 
at Born level, this
amounts to, approximately, one tenth of the total observable.
Therefore, in order to produce a visible effect, the modification of
the $Zb\bar b$ vertex should be, at the considered $q^2$ values,
exceptionally large. The second comment is that only 
the $Zb\bar b$ vertex
that \underline{does vary} with $q^2$ would be effective, since the
value of the vertex at $q^2=M^2_Z$ is automatically subtracted in our
procedure, and reabsorbed by the new theoretical input provided by the
$Zb\bar b$ partial width $\Gamma_b$ 
(whose value is now in essential agreement
with the SM, so that no sources of theoretical troubles are introduced
bu its use in our approach). In order to make an effect in $\sigma_5$
at the considered $q^2$ values, a TC model should therefore generate a
``huge" effect in the $Zb\bar b$ vertex 
that varies ``dramatically" with $q^2$
when one moves from $q^2=M^2_Z$
to higher values, which seems to us, at least for the preliminary LEP2
case, rather peculiar. We believe that this possibility might be
examined, in particular considering the specific observable
$\sigma_b(q^2)$, whose
future realistic experimentally accuracy remains to be fully
understood, as a special case whenever a theoretical model that meets
the previous requested would (will) be proposed.

\section{Concluding remarks}

\hspace{0.7cm}In this paper we have considered 
the information that can be obtained about
a strongly-interacting symmetry breaking sector from precision measurements
of four-fermion processes at LEP2 or a (500 GeV) NLC . Using a ``Z-peak''
subtracted approach to describe four-fermion processes, we have shown that
measurements of the cross section for muon production, the related
forward-backward asymmetry, and the total cross section for the production
of hadrons (except $t$'s) can place constraints on (or measure the effects
of) the lightest vector or axial resonances present in a strong
symmetry breaking sector. We estimate that such effects will be visible at LEP2
for resonances of masses up to approximately 350 GeV, and at a 500 GeV NLC
for resonances of masses up to approximately 800 GeV. Multiscale models, in
particular, predict the presence of light vector and axial mesons in this
mass range and their effects could be probed by these measurements.

The four-fermion processes measured here provide a complementary
probe to the $e^+ e^- \to W^+ W^-$ measurement suggested by Barklow
\cite{barklow}. The four-fermion processes are sensitive only to effects in
the gauge-boson self energies and, because of the ``Z-peak'' subtraction,
are mostly sensitive to the lightest scale objects in the symmetry breaking
sector. Two-gauge boson production, on the other hand, is sensitive largely
to gauge-boson rescattering effects which are likely to be dominated by
whatever sector of the model is responsible for the bulk of electroweak
symmetry breaking.

A final remark is that we did not consider, in the discussion of NLC
possibilities, that represented by the availability of longitudinally
polarized lepton beams. That would produce an enrichment of
experimental measurements, fully discussed in a previous paper
\cite{R26}, and
would lead to a definite improvement of bounds and of effects. The
reason why we did not include the discussion here is that, to our
knowledge, a fully rigorous discussion of the QED radiation effects for
these longitudinal polarization observables at NLC is still missing 
\cite{R27}.
The problem is being considered at the moment, and work along that
direction is already in progress.

{\bf \underline{Acknowledgments}}\par
C.V. wishes to thank the Department of Physique
Math\'ematique et Th\'eorique of Montpellier University and the
Department of Physics of Boston University for their warm hospitality
during the preparation of this paper. R.S.C. wishes to thank
the Department of Physics of the University of Lecce for its
hospitality during the inception and completion of this work.

\newpage

\vspace{1cm}
\begin{center}
{\bf Table A1:Standard Model values for the observables $\sigma_\mu$,
$A_{FB,\mu}$, and $\sigma_5$\\ for $\sqrt{q^2}$ between
150 and 190 GeV}\\
For each energy, the upper row shows the predictions
of the ``Z-subtracted'' approach $(S)$ and the lower row the predictions
of the complete calculation as encoded in the program TOPAZ0
$(T)$, ref.\cite{R21}.

\vspace{1cm}
\begin{tabular}{|c|c|c|c|} \hline
\multicolumn{1}{|c|}{$\sqrt{q^2}$ (GeV)}&
\multicolumn{1}{|c|}{${\sigma_{\mu}}^{(S)}_{(T)}$} &
\multicolumn{1}{|c|}{${A_{FB,\mu}}^{(S)}_{(T)}$ }&
\multicolumn{1}{|c|}{${\sigma_5}^{(S)}_{(T)}$} 
 \\[0.1cm] \hline
150& 5.842&0.635&45.176\\
&5.807&0.635&44.901\\ \hline
161&4.953&0.607&36.384\\ 
&4.876&0.605&35.547\\ \hline
165&4.661&0.600&33.579\\ 
&4.599&0.598&32.945\\ \hline
175&4.046&0.583&27.968\\
&4.013&0.583&27.710\\ \hline
180&3.788&0.576&25.729\\
&3.767&0.576&25.602\\ \hline
190&3.346&0.563&22.056\\ 
&3.340&0.564&22.115\\[0.1cm] \hline
\end{tabular}
\end{center}

\newpage

\begin{center}

{\large \bf Figure captions}
\end{center}
\vspace{0.5cm}

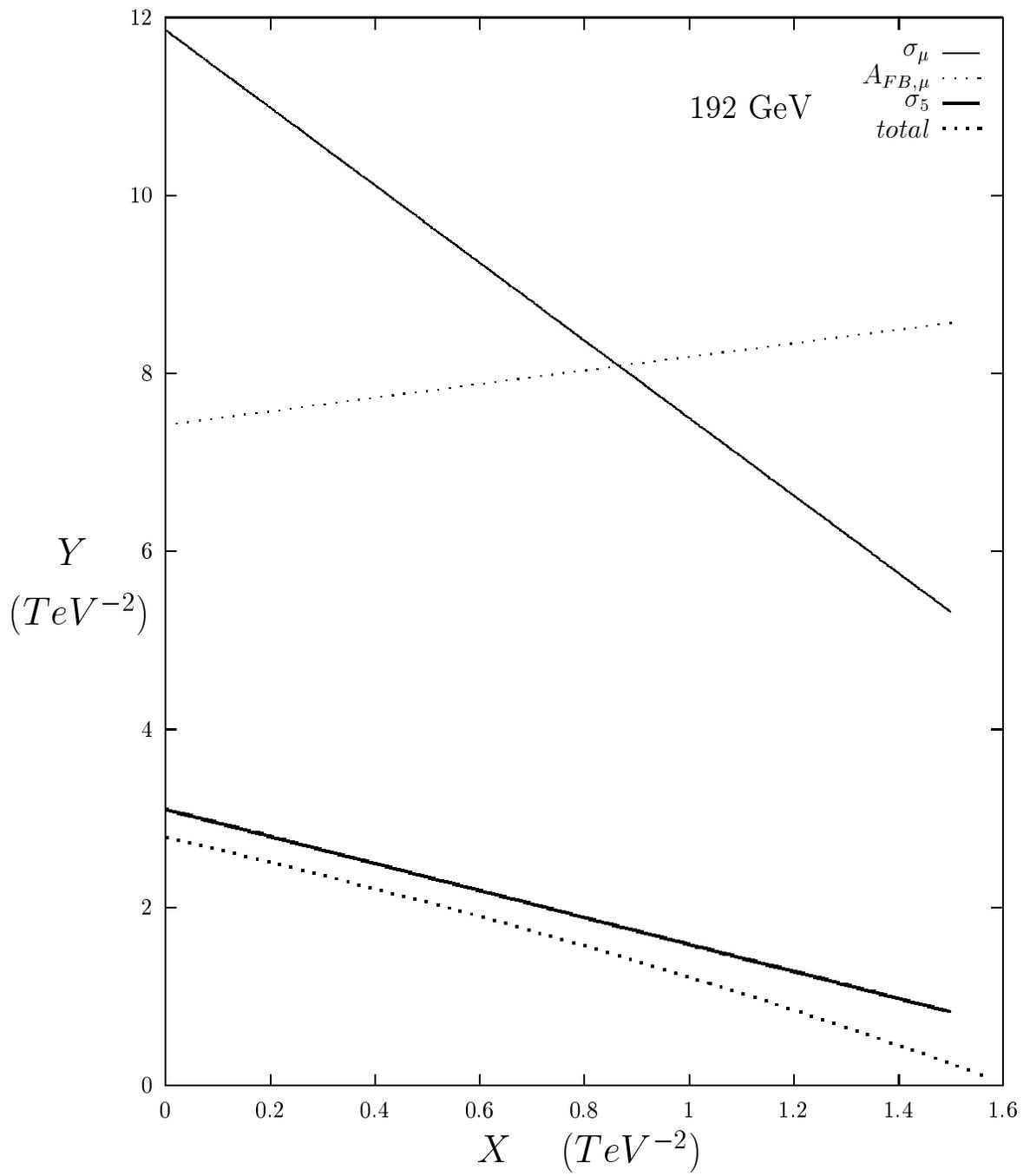
{\bf Fig.1} Limits (at $2\sigma$) on light TC parameters $X$,$Y$ at
LEP2(192 GeV), from $\sigma_{\mu}$
,$ A_{FB,\mu}$,
$\sigma_5$, and their
quadratic combinations (the observability domain lies above the
lines).\\

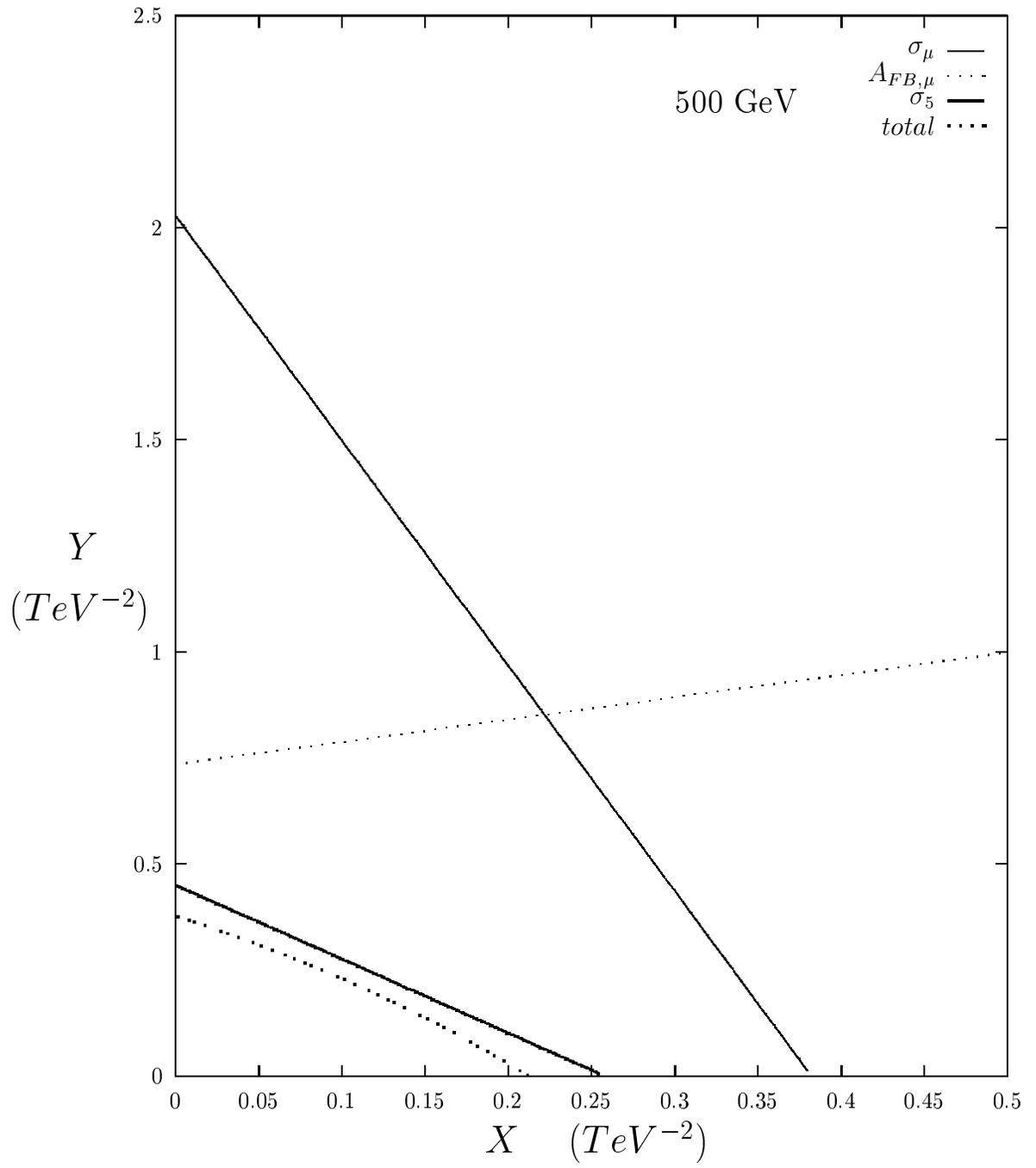
{\bf Fig.2} Limits (at $2\sigma$) on light TC parameters $X$,$Y$ at
NLC(500 GeV), same captions as in Fig.1. \\

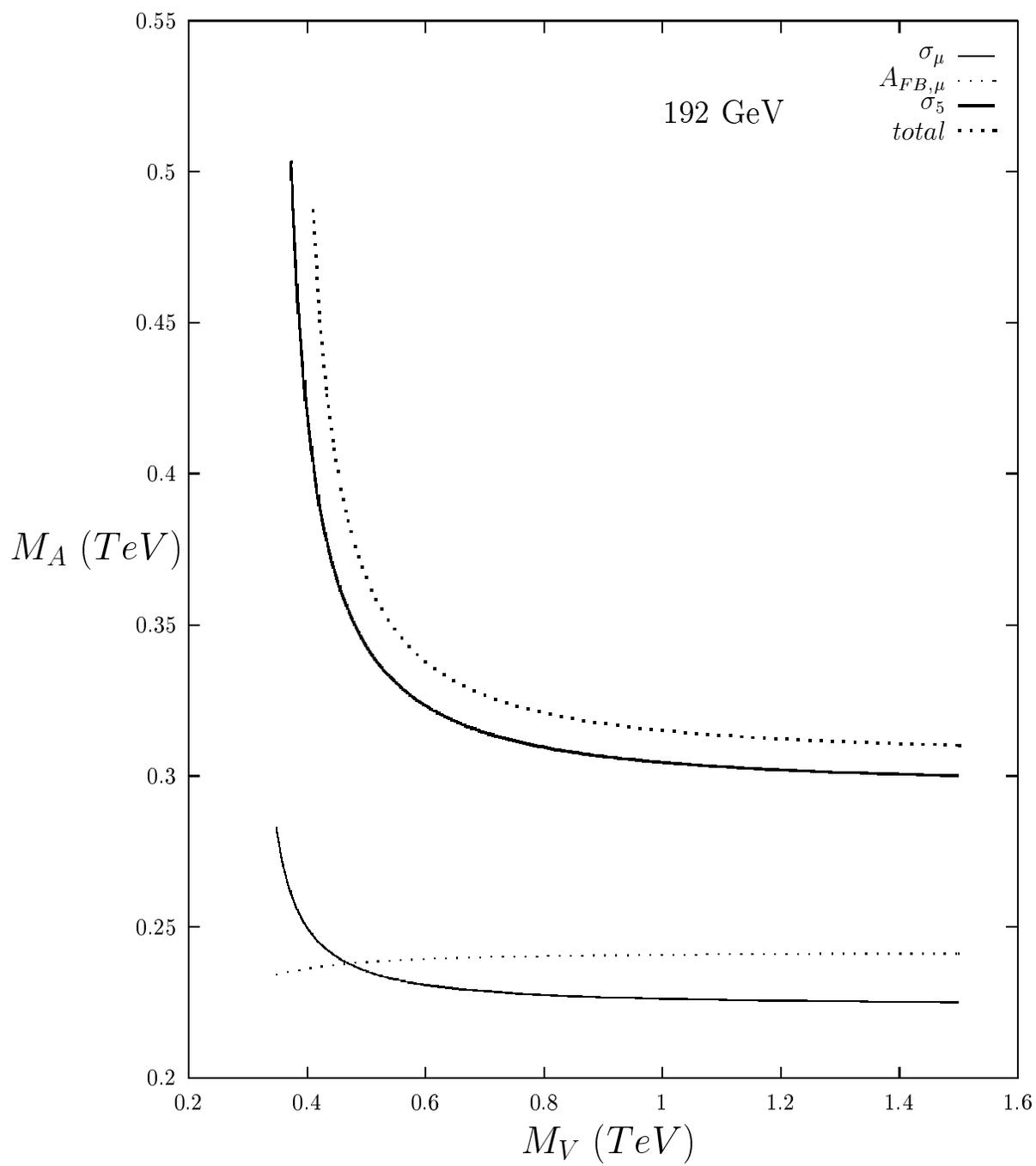
{\bf Fig.3} Limits (at $2\sigma$) on light TC masses $M_V$, $M_A$, in the
$\alpha=2$, $\beta=1$ model, at
LEP2(192 GeV), same captions as in Fig.1 but the observability  domain 
region lies below the lines.\\

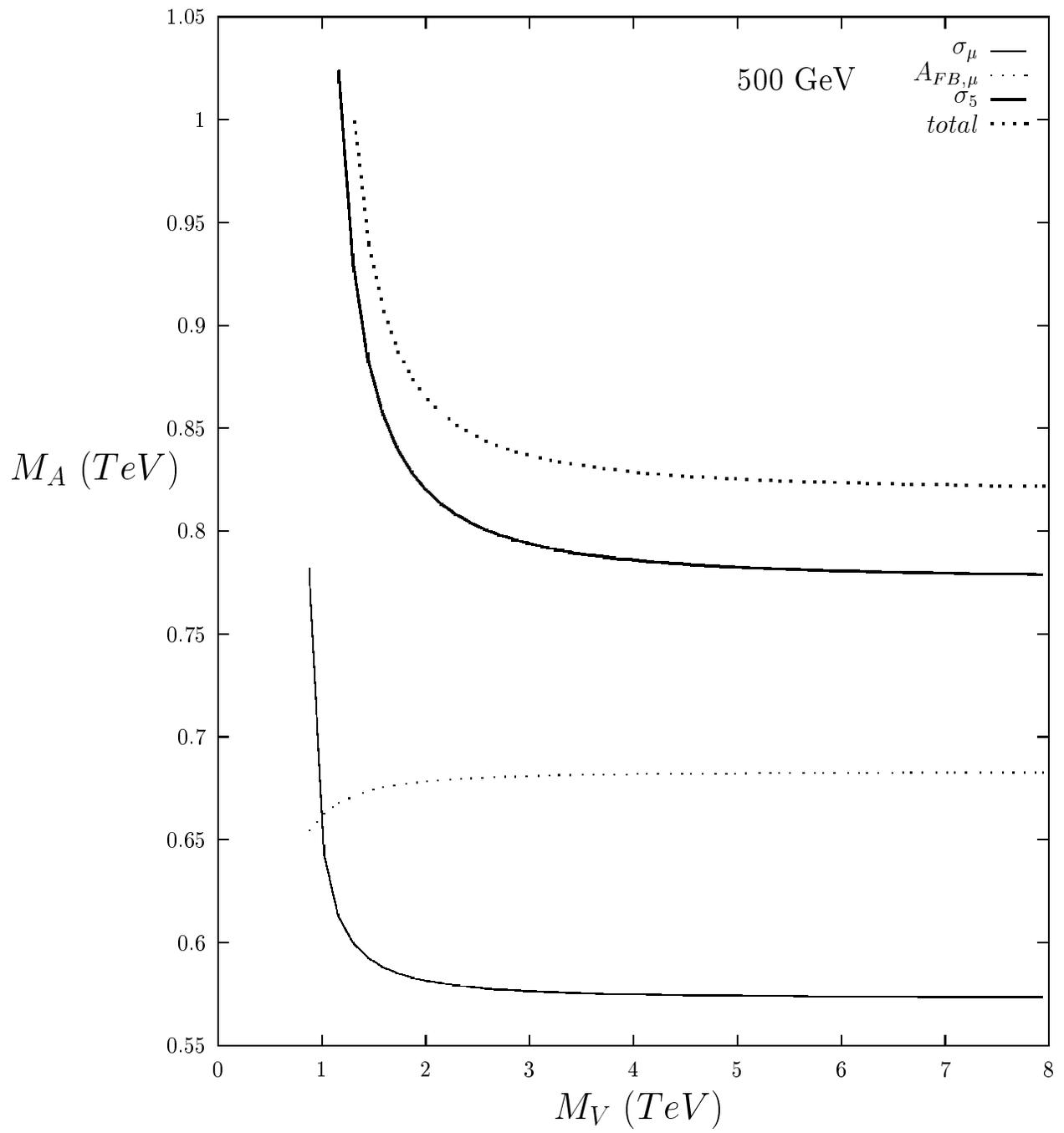
{\bf Fig.4} Limits (at $2\sigma$) on light TC masses $M_V$, $M_A$, in the
$\alpha=2$, $\beta=1$ model, at
NLC(500 GeV), same captions as in Fig.3.\\

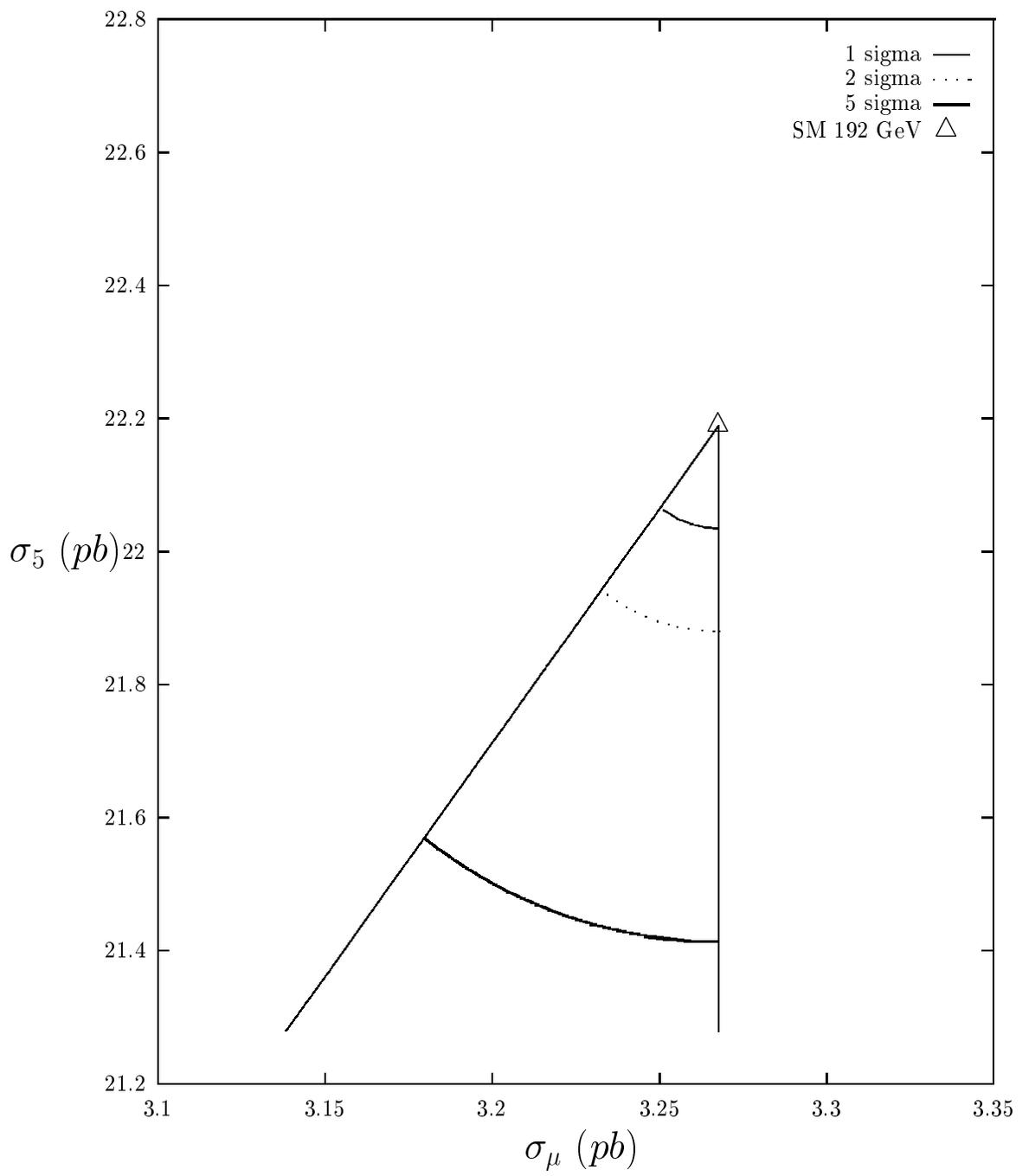
{\bf Fig.5} Allowed domain for light TC effects in the
($\sigma_5$, $\sigma_{\mu}$) plane, at LEP2(192 GeV). The three curves
correspond to 1,2,5 standard deviations, the observability 
domain lies inside the triangle.\\

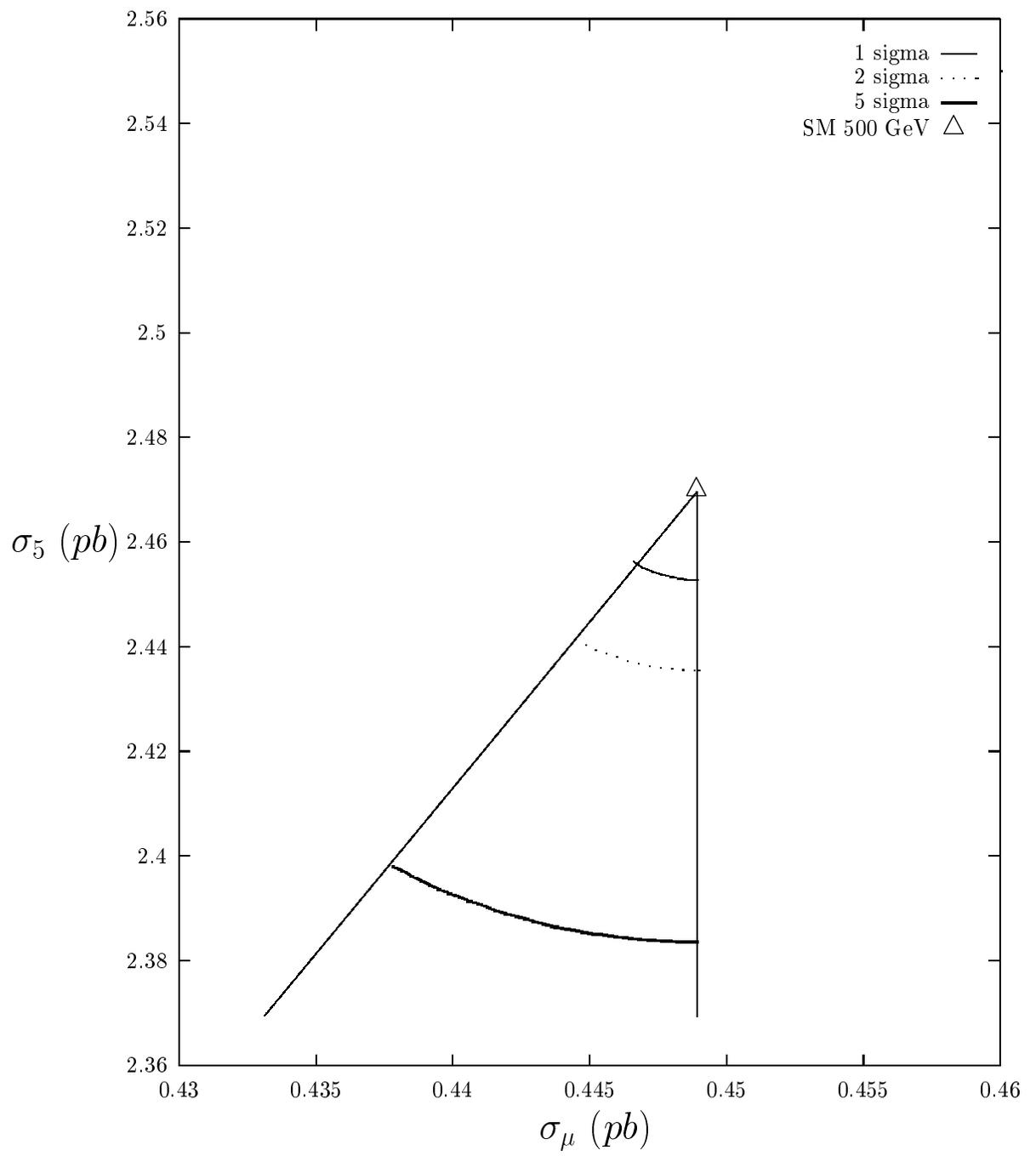
{\bf Fig.6} Allowed domain for light TC effects in the
($\sigma_5$, $\sigma_{\mu}$) plane, at NLC(500 GeV), same captions
as in Fig.5.

\end{document}